\documentstyle[psfig,12pt]{article}
\def\be{\begin{equation}}       
\def\ee{\end{equation}}
\def\bear{\be\begin{array}}      
\def\eear{\end{array}\ee}
\def\bea{\begin{eqnarray}}
\def\eea{\end{eqnarray}}

\include{mat}
\def\21{$SU(2) \ot U(1)$}
\def\ot{\otimes}
\def\ie{{\it i.e.}}

\def\quarter{{\textstyle{1 \over 4}}}

\def\eighth{{\textstyle{1 \over 8}}}
\def\bold#1{\setbox0=\hbox{$#1$}
     \kern-.025em\copy0\kern-\wd0
     \kern.05em\copy0\kern-\wd0
     \kern-.025em\raise.0433em\box0 }
\begin{document}
\begin{titlepage}
\begin{flushright}
IFIC/98-11\\
FTUV/98-11\\
hep-ph/9802407\\
February 1998
\end{flushright}
\vspace*{5mm}
\begin{center} 
{\Large \bf Bilinear R--Parity Violation${}^{\dag}$}\\[15mm]
{\large Marco Aurelio D\'\i az} \\
\hspace{3cm}\\
{\small Departamento de F\'\i sica Te\'orica, IFIC-CSIC, 
Universidad de Valencia}\\ 
{\small Burjassot, Valencia 46100, Spain}
\end{center}
\vspace{5mm}
\begin{abstract}

We review some of the main features of Bilinear R--Parity Violation
(BRpV), defined by a quadratic term in the superpotential which mixes 
lepton and Higgs superfields and is proportional to a mass parameter 
$\epsilon$. We show how large values of $\epsilon$ can induce a small 
neutrino mass without fine-tunning. We mention the effect on the mass 
of the lightest Higgs boson. Finally we report on the effect of 
BRpV on gauge and Yukawa unification, showing that bottom--tau 
unification can be achieved at any value of $\tan\beta$.

\end{abstract}

\vskip 5.cm
\noindent ${}^{\dag}$Talk given at the International Workshop
``Beyond the Standard Model: From Theory to Experiment'', 
13--17 October 1997, Valencia, Spain.

\end{titlepage}

\setcounter{page}{1}

The Standard Model (SM) works well in describing the phenomenology of the
strong and electroweak interactions of the known particles. For this
reason, the motivations for studying supersymmetric extensions of the 
SM are mostly theoretical. The only experimental indication that favors
the Minimal Supersymmetric Standard Model (MSSM) \cite{MSSM} in 
comparison with the SM is the unification of gauge couplings at some high 
scale $M_{GUT}$ \cite{gaugeUnif,gaugUnifRecent}. 

Supersymmetry \cite{SUSY} is the only known way of unifying non--trivially 
the space-time symmetries of the Poincar\`e group with some other internal 
symmetry. This symmetry relates bosons with fermions, and at the same 
time affects the notion of space-time itself by introducing anti-commuting 
coordinates which extends the Minkowsky space into a superspace.
Superfields are functions of superspace coordinates and the MSSM is
constructed with vector and scalar superfields. Vector superfields 
$\widehat V$ contains a spin--1 gauge boson $v_{\mu}$ and a fermionic 
partner $\lambda$ (for example the photon and the photino in 
supersymmetric electrodynamics). Scalar superfields $\widehat\Phi$ 
contains a scalar boson $\phi$ and a fermionic partner $\psi$ (for example 
Higgs bosons and higgsinos). The superpotential is a cubic polynomial 
function of superfields.

It is costumary to assign to each component field an R--Parity defined by 
$R_p=(-1)^{3B+L+2S}$, where $B$ is the barion number, $L$ is the lepton
number and $S$ is the spin. In this way, quarks, leptons and Higgs bosons
are R--Parity even, and the supersymmetric particles are R--Parity odd.
If R--Parity is conserved, then supersymmetric particles are produced in
pairs in the laboratory. In addition, the lightest supersymmetric particle
(LSP, the lightest neutralino) is stable. 

On the contrary, if R--Parity is not conserved then supersymmetric 
particles can be single produced, and the LSP decays into standard quarks 
and leptons. Furthermore, the LSP needs not to be the lightest neutralino.
Possible terms in the superpotential which violate R--Parity are
\begin{equation} 
W_{R_p\!\!\!\!\!\!/}=\lambda''_{ijk}\widehat U_i\widehat D_j\widehat D_k+
\varepsilon_{ab}\left[
 \lambda'_{ijk}\widehat L_i^a\widehat Q_j^b\widehat D_k
+\lambda_{ijk}\widehat L_i^a\widehat L_j^b\widehat R_k
+\epsilon_i\widehat L_i^a\widehat H_2^b\right]\,,
\label{WnotRp}
\end{equation}
Trilinear R--Parity Violation (TRpV) corresponds to the first three terms 
and, considering that each of the generation indices $i,j,k$ run from 1 
to 3, they involve a very large number of arbitrary parameters. The only 
practical way to study TRpV is to consider one or two $\lambda$'s 
different from zero at a time.

The fourth term in eq.~(\ref{WnotRp}) corresponds to Bilinear R--Parity 
Violation (BRpV) \cite{e3others,BRpVtalk}, and involves only three extra 
parameters, one $\epsilon_i$ for each generation. The $\epsilon_i$ terms 
also violate lepton number in the $i$th generation respectively. Models 
where R--Parity is spontaneously broken \cite{SRpSB} through vacuum 
expectation values (vev) of right handed sneutrinos 
$\langle{\tilde\nu^c}\rangle=v_R\neq0$ generate BRpV (and not 
TRpV)\footnote{
Of course, this is true in the original basis. If we rotate the Higgs 
and Lepton superfields then TRpV terms are generated, as explained later.
}. 
The $\epsilon_i$ parameters are then equal to some Yukawa coupling 
times $v_R$. Motivated by spontaneously broken R--Parity, we introduce 
explicitly BRpV in the MSSM superpotential and review the most 
important features of this model.

For simplicity we take from now on $\epsilon_1=\epsilon_2=0$, in this way, 
only tau--lepton number is violated. In this case, considering only the 
third generation, the MSSM--BRpV has the following superpotential
\begin{equation} 
W=\varepsilon_{ab}\left[
 h_t\widehat Q_3^a\widehat U_3\widehat H_2^b
+h_b\widehat Q_3^b\widehat D_3\widehat H_1^a
+h_{\tau}\widehat L_3^b\widehat R_3\widehat H_1^a
-\mu\widehat H_1^a\widehat H_2^b
+\epsilon_3\widehat L_3^a\widehat H_2^b\right]\,,
\label{eq:Wsuppot}
\end{equation}
where the first four terms correspond to the MSSM. The last term violates
tau--lepton number as well as R--Parity. 

The presence of the $\epsilon$ term in the superpotential implies that
the tadpole equation for the tau sneutrino is non--trivial, \ie, the
vacuum expectation value 
$\left\langle\tilde\nu_{\tau}\right\rangle=v_3/\sqrt{2}$ 
is non--zero. This in turn generates more R--parity and tau lepton number
violating terms which, in particular, induce a tau neutrino mass as we
will see later.

By looking at the last two terms in the superpotential an immediate 
question arises. Can the BRpV term be rotated away from the 
superpotential? and consequently, is the $\epsilon$ term physical? 
Indeed, consider the following rotation of the superfields \cite{HallSuzuki}
\begin{equation}
\widehat H_1'={{\mu\widehat H_1-\epsilon_3\widehat L_3}\over{
\sqrt{\mu^2+\epsilon_3^2}}}\,,\qquad
\widehat L_3'={{\epsilon_3\widehat H_1+\mu\widehat L_3}\over{
\sqrt{\mu^2+\epsilon_3^2}}}\,.
\label{eq:rotation}
\end{equation}
In the new basis the $\epsilon$ term disappears from the superpotential,
nevertheless, R--Parity is reintroduced in the form of TRpV. The 
superpotential in the new basis is
\begin{equation} 
W=h_t\widehat Q_3\widehat U_3\widehat H_2
+h_b{{\mu}\over{\mu'}}\widehat Q_3\widehat D_3\widehat H'_1
+h_{\tau}\widehat L'_3\widehat R_3\widehat H'_1
-\mu'\widehat H'_1\widehat H_2
+h_b{{\epsilon_3}\over{\mu'}}\widehat Q_3\widehat D_3\widehat L'_3
\,,\label{WsuppotP}
\end{equation}
where $\mu'^2=\mu^2+\epsilon_3^2$. The first four terms are MSSM looking
terms and the last term violates the R--Parity defined in the new basis.
Note the re-scaling in the bottom quark Yukawa term. Its presence
ensures that the same quark mass is obtained with the same Yukawa
coupling in the two basis. This re-scaling is non-trivial and has
important consequences in Yukawa unification, as shown later.

As we know, supersymmetry must is broken and this is parametrized by 
soft supersymmetry breaking terms. The soft terms which play an 
important role in BRpV are the following
\begin{equation}
V_{soft}=m_{H_1}^2|H_1|^2+M_{L_3}^2|\widetilde L_3|^2
-\left[B\mu H_1H_2-B_2\epsilon_3\widetilde L_3H_2+h.c.\right]+...
\label{SoftUnrot}
\end{equation}
where $m_{H_1}^2$ and $M_{L_3}^2$ are the soft masses corresponding to 
the fields $H_1$ and $\widetilde L_3$ respectively, and $B$ and $B_2$ 
are the bilinear soft mass parameters associated to the next-to-last 
and last terms in the superpotential in eq.~(\ref{eq:Wsuppot}). It is
clear, for example, that Higgs vacuum expectation values 
$\left\langle H_i\right\rangle=v_i/\sqrt{2}$ induce a non-trivial
tadpole equation and a non-zero vev for the sneutrino through the 
$B_2$ term in eq.~(\ref{SoftUnrot}).

The soft terms in the rotated basis are given by
\begin{eqnarray}
V_{soft}&=&
{{m_{H_1}^2\mu^2+M_{L_3}^2\epsilon_3^2}\over{\mu'^2}}|H'_1|^2
+{{m_{H_1}^2\epsilon_3^2+M_{L_3}^2\mu^2}\over{\mu'^2}}|\widetilde L'_3|^2-
\bigg[{{B\mu^2+B_2\epsilon_3^2}\over{\mu'}}H'_1H_2
\nonumber\\&&
-{{\epsilon_3\mu}\over{\mu'^2}}(m_{H_1}^2-M_{L_3}^2)\widetilde L'_3H'_1
-{{\epsilon_3\mu}\over{\mu'}}(B_2-B)\widetilde L'_3H_2+h.c.\bigg]+...
\label{SoftRotated}
\end{eqnarray}
The first three terms are MSSM like terms equivalent to the first three
terms in eq.~(\ref{SoftUnrot}). In fact, in analogy with the MSSM, the 
coefficients of $|H'_1|^2$ and $|\widetilde L'_3|^2$ could be defined in
the rotated basis as the soft masses $m'^2_{H_1}$ and $M'^2_{L_3}$ 
respectively, and the coefficient of $H'_1H_2$ would be the new bilinear 
soft term $B'\mu'$. The last two terms violate R--Parity and 
tau lepton number, and are equivalent to the last term in 
eq.~(\ref{SoftUnrot}), \ie, they induce a non-zero vev for the tau 
sneutrino field in the rotated basis 
$\left\langle\tilde\nu'_{\tau}\right\rangle=v'_3/\sqrt{2}$ \cite{Basis}.

Vacuum expectation values are calculated by minimizing the scalar 
potential, or equivalently, by imposing that the tadpoles are equal to
zero. The linear terms of the scalar potential are
$V_{linear}=t_1\chi^0_1+t_2\chi^0_2+t_3\tilde\nu^R_{\tau}$,
where $\chi^0_i=\sqrt{2}Re(H^i_i)-v_i$ and 
$\tilde\nu^R_{\tau}=\sqrt{2}Re(\tilde\nu_{\tau})-v_3$. The $t_i$ are the 
tree level tadpoles and they are equal to zero at the minimum. In the
original basis the tadpole equations are
\begin{eqnarray}
t_1&=&(m_{H_1}^2+\mu^2)v_1-B\mu v_2-\mu\epsilon_3v_3+
\eighth(g^2+g'^2)v_1(v_1^2-v_2^2+v_3^2)=0\,,
\nonumber \\
t_2&=&(m_{H_2}^2+\mu^2+\epsilon_3^2)v_2-B\mu v_1+B_2\epsilon_3v_3-
\eighth(g^2+g'^2)v_2(v_1^2-v_2^2+v_3^2)=0\,,
\nonumber \\
t_3&=&(M_{L_3}^2+\epsilon_3^2)v_3-\mu\epsilon_3v_1+B_2\epsilon_3v_2+
\eighth(g^2+g'^2)v_3(v_1^2-v_2^2+v_3^2)=0\,.
\label{eq:tadpoles}
\end{eqnarray}
The first two tadpole equations reduce to the MSSM minimization conditions
after taking the MSSM limit $\epsilon_3=v_3=0$, and in this case, the 
third tadpole equation is satisfied trivially. Note that $\epsilon_3=0$
implies two solutions for $v_3$ from the third tadpole in 
eq.~(\ref{eq:tadpoles}), from which only $v_3=0$ is viable because the 
second solution implies the existence of a massless pseudoscalar.

The first two tadpole equations in the rotated basis are
\begin{eqnarray}
t'_1&=&\mu'^2v'_1+{{m_{H_1}^2\mu^2+M_{L_3}^2\epsilon_3^2}\over{\mu'^2}}
v'_1-{{B\mu^2+B_2\epsilon_3^2}\over{\mu'}}v_2+
(m_{H_1}^2-M_{L_3}^2){{\epsilon_3\mu}\over{\mu'^2}}v'_3
\nonumber\\&&
+{\textstyle{1\over8}}(g^2+g'^2)v'_1(v'^2_1-v_2^2+v'^2_3)=0
\label{UnoTadpoleRot}\\ \nonumber\\
t'_2&=&\mu'^2v_2+m_{H_2}^2v_2-{{B\mu^2+B_2\epsilon_3^2}\over{\mu'}}v'_1
+(B_2-B){{\epsilon_3\mu}\over{\mu'}}v'_3
\nonumber\\&&
-{\textstyle{1\over8}}(g^2+g'^2)v_2(v'^2_1-v_2^2+v'^2_3)=0
\label{TwoTadpoleRot}
\end{eqnarray}
where $\left\langle H'_1\right\rangle=v'_1/\sqrt{2}$ and 
$\langle\widetilde L'_3\rangle=v'_3/\sqrt{2}$, and the following relations
hold $v'_1=(\mu v_1-\epsilon_3v_3)/\mu'$ and 
$v'_3=(\epsilon_3v_1+\mu v_3)/\mu'$, as suggested by 
eq.~(\ref{eq:rotation}). These two tadpole equations
resemble the MSSM minimization conditions when we set $v'_3=0$. The third
tadpole equation is
\begin{eqnarray}
t'_3&=&(m_{H_1}^2-M_{L_3}^2){{\epsilon_3\mu}\over{\mu'^2}}v'_1
+(B_2-B){{\epsilon_3\mu}\over{\mu'}}v_2
+{{m_{H_1}^2\epsilon_3^2+M_{L_3}^2\mu^2}\over{\mu'^2}}v'_3
\nonumber\\&&
+{\textstyle{1\over8}}(g^2+g'^2)v'_3(v'^2_1-v_2^2+v'^2_3)=0
\label{tadpoleiii}
\end{eqnarray}
In this equation we observe that $v'_3=0$ if 
$\Delta m^2\equiv m_{H_1}^2-M_{L_3}^2=0$ and $\Delta B\equiv B_2-B=0$ at the
weak scale, which is not true in general. In supergravity models with 
universality of scalar soft masses and bilinear mass parameters we
have $\Delta m^2=0$ and $\Delta B=0$ at the unification scale 
$M_{GUT}\approx 10^{16}$ GeV, but radiative corrections spoil this 
degeneracy. In the approximation where $\Delta m^2$ and $\Delta B$ are small
we find that $v'_3$ is also small and in first approximation given by
\begin{equation}
v'_3\approx -{{\epsilon_3\mu}\over{\mu'^2m_{\tilde\nu^0_{\tau}}^2}}
\left(v'_1\Delta m^2+\mu'v_2\Delta B\right)
\label{App_v3p}
\end{equation}
where we have introduced
\begin{equation}
m_{\tilde\nu^0_{\tau}}^2\equiv {{m_{H_1}^2\epsilon_3^2+M_{L_3}^2\mu^2}
\over{\mu'^2}}+{\textstyle{1\over8}}(g^2+g'^2)(v'^2_1-v_2^2)
\label{sneumassMSSM}
\end{equation}
which reduces to the tau sneutrino mass in the MSSM when we set 
$\epsilon_3=0$.

\begin{figure}
\centerline{\protect\hbox{\psfig{file=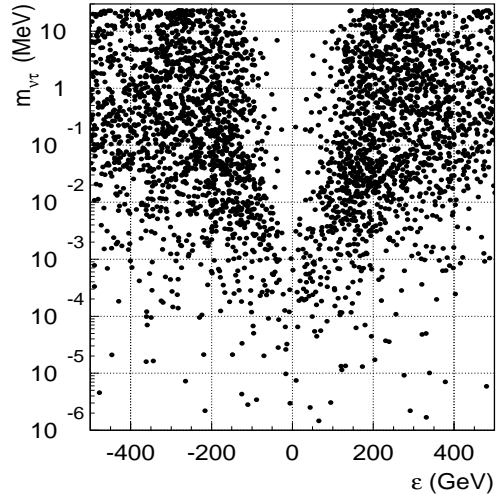,height=7cm,width=0.43\textwidth}}}
\caption{Tau neutrino mass as a function of the R--Parity violating 
parameter $\epsilon_3$.}
\label{mnt_e3_ev}
\end{figure}

As a consequence of tau lepton number and BRpV terms, characterized by 
the parameters $\epsilon_3$ and $v_3$, a mixing between neutralinos and 
the tau neutrino is generated. This implies that the tau neutrino acquires 
a mass $m_{\nu_{\tau}}$. In the original basis, where 
$(\psi^0)^T= 
(-i\lambda',-i\lambda^3,\widetilde{H}_1^1,\widetilde{H}_2^2,\nu_{\tau})$,
the scalar potential contains the following mass terms
\begin{equation} 
{\cal L}_m=-\frac 12(\psi^0)^T{\bold M}_N\psi^0+h.c.   
\label{eq:NeuMLag} 
\end{equation} 
where the neutralino/neutrino mass matrix is 
\begin{equation} 
{\bold M}_N=\left[  
\begin{array}{ccccc}  
M^{\prime } & 0 & -\frac 12g^{\prime }v_1 & \frac 12g^{\prime }v_2 & -\frac  
12g^{\prime }v_3 \\   
0 & M & \frac 12gv_1 & -\frac 12gv_2 & \frac 12gv_3 \\   
-\frac 12g^{\prime }v_1 & \frac 12gv_1 & 0 & -\mu  & 0 \\   
\frac 12g^{\prime }v_2 & -\frac 12gv_2 & -\mu  & 0 & \epsilon _3 \\   
-\frac 12g^{\prime }v_3 & \frac 12gv_3 & 0 & \epsilon _3 & 0  
\end{array}  
\right] 
\label{eq:NeuM5x5} 
\end{equation} 
Here $M$ and $M'$ are the $SU(2)$ and $U(1)$ gaugino masses. It can be 
seen from eq.~(\ref{eq:NeuM5x5}) that mixings between tau neutrino and 
neutralinos are proportional to $\epsilon_3$ and $v_3$. Naively one could
think that, due to the strong experimental constraint on the tau neutrino 
mass, the parameters $\epsilon_3$ and $v_3$ should be small compared
with $m_Z$. This is not the case, and from Fig.~\ref{mnt_e3_ev} we 
observe that $|\epsilon_3|$ can be as large as 400 GeV! 

Indeed, to make Fig.~\ref{mnt_e3_ev} we have embedded the MSSM--BRpV
model into supergravity \cite{epsrad}, with universality of scalar 
($m_0$), gaugino ($M_{1/2}$), bilinear ($B$), and trilinear ($A$) soft 
mass parameters at the unification scale $M_X\approx10^{16}$ GeV. We have
imposed the radiative breaking of the electroweak symmetry by minimizing
the scalar potential with the aid of one--loop tadpole equations. We
have made a scan over the parameter space, including the BRpV parameters 
$\epsilon_3$ and $v_3$. Points that satisfy the constraint 
$m_{\nu_{\tau}}<30$ MeV are kept (we also impose that the 
supersymmetric particles are not too light). 

We observe from Fig.~\ref{mnt_e3_ev} that it is easy to satisfy the 
constraint on the tau neutrino mass, and even $m_{\nu_{\tau}}$ of 
the order of 1 eV can be achieved. The central region where $\epsilon_3$ 
is close to zero is less populated at high values of 
$m_{\nu_{\tau}}$ because in this case we are closer to the MSSM, 
where the neutrinos are massless.

\begin{figure}
\centerline{\protect\hbox{\psfig{file=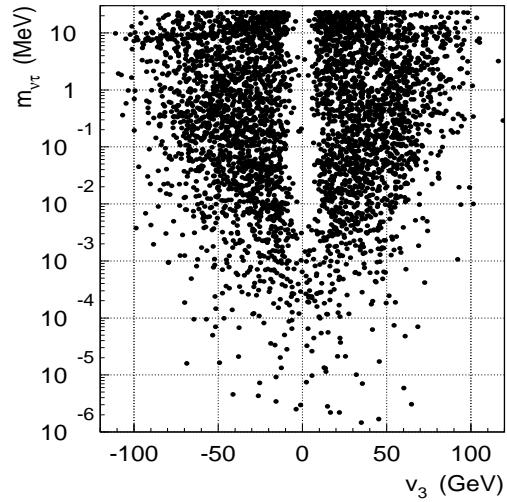,height=7cm,width=0.43\textwidth}}}
\caption{Tau neutrino mass as a function of the tau sneutrino vacuum 
expectation value $v_3$ in the original basis.}
\label{mnt_v3}
\end{figure}
Similarly, in Fig.~\ref{mnt_v3} we plot $m_{\nu_{\tau}}$ as a 
function of the vacuum expectation value of the sneutrino $v_3$. For 
the same reason we already mention, the central region at high values
of $m_{\nu_{\tau}}$ is less populated. In this figure we observe
that the BRpV parameter $v_3$ is not necessarily small, and that $|v_3|$ 
can be as high as 100 GeV. The value of $|v_3|$ cannot be as high as 
$\epsilon_3$ because the sneutrino vev contributes also to the 
$W$--boson mass according to 
$m_W^2=\quarter(g^2+g'^2)(v_1^2+v_2^2+v_3^2)$.

Considering that the mass terms which mix the neutrino with the 
neutralinos are proportional to $\epsilon_3$ and $v_3$, an obvious 
question arises: how can we get a small neutrino mass? The answer lies
in the fact that the induced neutrino mass satisfy
$m_{\nu_{\tau}}\sim(\epsilon_3v_1+\mu v_3)^2$, and this last
combination is what needs to be small. Indeed, as we will see below, 
in models with universality of scalar and bilinear soft mass parameters,
the combination $(\epsilon_3v_1+\mu v_3)$ is radiatively induced, and
therefore, naturally small.

\begin{figure}
\centerline{\protect\hbox{\psfig{file=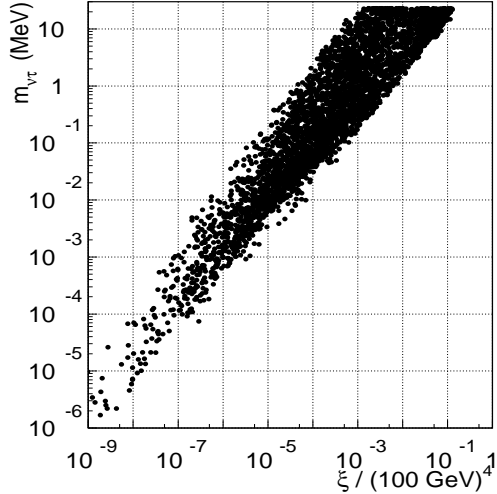,height=7cm,width=0.43\textwidth}}}
\caption{Tau neutrino mass as a function of  
$\xi\equiv(\epsilon_3v_1+\mu v_3)^2$, which is related to the v.e.v. of the
tau sneutrino in the rotated basis through $\xi=(\mu'v_3')^2$.}
\label{mnt_xi_ev}
\end{figure}
In Fig.~\ref{mnt_xi_ev} we have the dependence of the tau neutrino mass
$m_{\nu_{\tau}}$ as a function of the parameter 
$\xi\equiv(\epsilon_3v_1+\mu v_3)^2=(\mu'v'_3)^2$. We see a clear
correlation between $m_{\nu_{\tau}}$ and $v'_3$. The parameter 
$|v'_3|$ takes a maximum value of the order of 10 GeV.

The neutralino--neutrino mass matrix in the rotated basis, analogous to
eq.~(\ref{eq:NeuM5x5}), is ${\bold M}'_N=R({\bold M}_N)$, where the 
rotation $R$ is defined by eq.~(\ref{eq:rotation}) or, equivalently, 
by the substitution
$(v_1,v_3,\epsilon_3,\mu)\longrightarrow(v'_1,v'_3,0,\mu')$. In this
basis the $\epsilon$ term is not present, and the only source of
mixing responsible for the neutrino mass is the vev $v'_3$. In first
approximation, valid when $v'_3$ is small, we get
\begin{equation}
m_{\nu_{\tau}}\approx-{{(g^2M+g'^2M')\mu'^2v'^2_3}\over{
4MM'\mu'^2-2(g^2M+g'^2M')v'_1v_2\mu'}}
\label{mNeutrinoApp}
\end{equation}
On the other hand, considering the renormalization group equations for
the soft mass parameters $m^2_{H_1}$, $m^2_{L_3}$, $B$, and $B_2$, which
solved in first approximation give us
\begin{eqnarray}
m_{H_1}^2-M_{L_3}^2&\approx&-{{3h_b^2}\over{8\pi^2}}
\left(m_{H_1}^2+M_Q^2+M_D^2+A_D^2\right)\ln{{M_{GUT}}\over{m_Z}}
\nonumber\\
B_2-B&\approx&{{3h_b^2}\over{8\pi^2}}A_D\ln{{M_{GUT}}\over{m_Z}}
\label{m2BDiff}
\end{eqnarray}
we can show, using eq.~(\ref{App_v3p}), that the tau neutrino mass is
radiatively generated and given by
\begin{equation}
m_{\nu_{\tau}}\approx{{
\left[\mu'v_2A_D-v'_1\left(m_{H_1}^2+M_Q^2+M_D^2+A_D^2\right)\right]^2
}\over{\Big[2v'_1v_2-4MM'\mu'/(g^2M+g'^2M')\Big]
\mu'm_{\tilde\nu^0_{\tau}}^2}}
\left({{\epsilon_3\mu}\over{\mu'^2}}\right)^2
\left({{3h_b^2}\over{8\pi^2}}\ln{{M_{GUT}}\over{m_Z}}\right)^2
\label{mNeutriApp2}
\end{equation}
and, therefore, naturally small. This mass can be further approximated by
\begin{equation}
m_{\nu_{\tau}}\approx{{m_Z^2}\over{M_{SUSY}}}
\left({{\epsilon_3}\over{M_{SUSY}}}\right)^2h_b^4\sim 1\,{\mathrm{KeV}}
\label{mNeutriApp3}
\end{equation}
where the 1 KeV was obtained in the case $M_{SUSY}\sim\epsilon_3\sim m_Z$
and $h_b\sim 10^{-2}$. An even lighter $\nu_{\tau}$ can be obtained if we 
increase $M_{SUSY}$ or decrease $\epsilon_3$, as can be seen from 
Fig.~\ref{mnt_xi_ev}, where neutrinos as light as $m_{\nu_{\tau}}\sim 1$ 
eV are shown \cite{epsrad}.

Another interesting feature of BRpV is that the neutral CP--even Higgs 
sector now mixes with the real part of the tau sneutrino forming a set
of three neutral CP--even scalars $S^0_i$, $i=1,2,3$. In the original 
basis, where $S^0=[\chi^0_1,\chi^0_2,\tilde\nu_{\tau}^R]$, the mass matrix 
is given by
\begin{eqnarray} 
&&{\bold M_{S^0}^2}= 
\label{eq:PseScalM}\\ \nonumber\\ 
&& \!\!\!
\left[\matrix{ 
B\mu{{v_2}\over{v_1}}+\quarter g_Z^2v_1^2+\mu\epsilon_3 
{{v_3}\over{v_1}}
& -B\mu-\quarter g^2_Zv_1v_2  
& -\mu\epsilon_3+\quarter g^2_Zv_1v_3  
\cr -B\mu-\quarter g^2_Zv_1v_2  
& \!\!\!\!\!\! B\mu{{v_1}\over{v_2}}+\quarter g^2_Zv_2^2-B_2\epsilon_3 
{{v_3}\over{v_2}}
& B_2\epsilon_3-\quarter g^2_Zv_2v_3  
\cr -\mu\epsilon_3+\quarter g^2_Zv_1v_3  
& B_2\epsilon_3-\quarter g^2_Zv_2v_3  
& \!\!\!\!\!\! \mu\epsilon_3{{v_1}\over{v_3}}
-B_2\epsilon_3{{v_2}\over{v_3}}+\quarter g^2_Zv_3^2
}\right] \nonumber 
\end{eqnarray} 
\begin{figure}
\centerline{\protect\hbox{\psfig{file=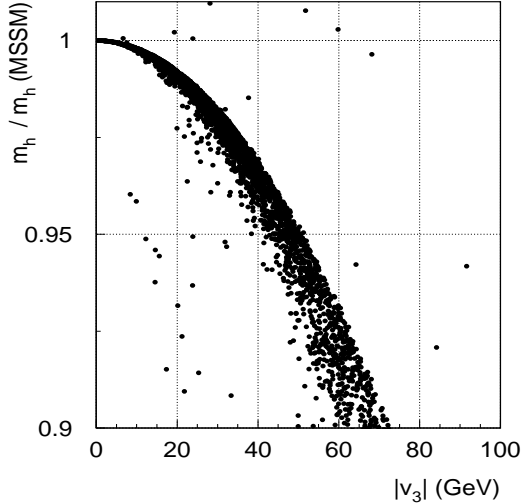,height=7cm,width=0.43\textwidth}}}
\caption{Ratio between the lightest CP-even neutral scalar mass in the 
$\epsilon$--model and the lightest CP--even Higgs mass in the MSSM, as a 
function of the tau sneutrino vacuum expectation value $v_3$.}
\label{ratio_v3}
\end{figure}
In the MSSM limit, where $\epsilon_3=v_3=0$, the mass matrix 
${\bold M_{S^0}^2}$ reduced to a $2\times2$ block corresponding to the
normal CP--even Higgs sector of the MSSM, and a decoupled tau sneutrino.
A similar effect occurs with the charged Higgs sector, which couples
to the stau sector forming a set of four charged scalars, one of them
being the unphysical Goldstone boson \cite{ChaStau}.

We have calculated the lightest CP--even Higgs mass in BRpV and compared it
with its mass in the MSSM and the result is plotted in Fig.~\ref{ratio_v3}.
We include only the largest radiative corrections proportional to $m_t^4$
\cite{mhRadCorr}. We observe that the lightest Higgs mass is in general 
decreased due to the mixing with the sneutrino and, of course, the effect
disappear as the BRpV parameter $|v_3|$ approaches to zero. In this case 
the Higgs $h$ have R--Parity violating decays because it can ``behaves''
as a tau sneutrino.

Similarly to the Higgs bosons, charginos mix with the tau lepton 
forming a set of three charged fermions $F_i^{\pm}$, $i=1,2,3$. In the 
original basis where $\psi^{+T}=(-i\lambda^+,\widetilde H_2^1,\tau_R^+)$ 
and $\psi^{-T}=(-i\lambda^-,\widetilde H_1^2,\tau_L^-)$, the charged 
fermion mass terms in the lagrangian are 
${\cal L}_m=-\psi^{-T}{\bold M_C}\psi^+$, with the mass matrix given by
\begin{equation} 
{\bold M_C}=\left[\matrix{ 
M & {\textstyle{1\over{\sqrt{2}}}}gv_2 & 0 \cr 
{\textstyle{1\over{\sqrt{2}}}}gv_1 & \mu &  
-{\textstyle{1\over{\sqrt{2}}}}h_{\tau}v_3 \cr 
{\textstyle{1\over{\sqrt{2}}}}gv_3 & -\epsilon_3 & 
{\textstyle{1\over{\sqrt{2}}}}h_{\tau}v_1}\right] 
\label{eq:ChaM6x6} 
\end{equation} 
As a result, the tau Yukawa coupling is not related to the tau mass by the
usual MSSM relation. On the contrary, $h_{\tau}$ depends now on the
parameters of the chargino sector $M$, $\mu$, and $\tan\beta$, as well as 
the BRpV parameters $\epsilon_3$ and $v_3$, through a formula given
in ref.~\cite{ChaStau}. In addition, the top and bottom quark Yukawa
couplings are related to the quark masses by
\begin{equation}
m_t = h_t {v\over{\sqrt2}} \sin \beta \sin \theta\,,\qquad
m_b = h_b {v\over{\sqrt2}} \cos \beta \sin \theta 
\end{equation}
where $v=246$ GeV and we have defined $\cos\theta\equiv v_3/v$.

\begin{figure}
\centerline{\protect\hbox{ 
\psfig{file=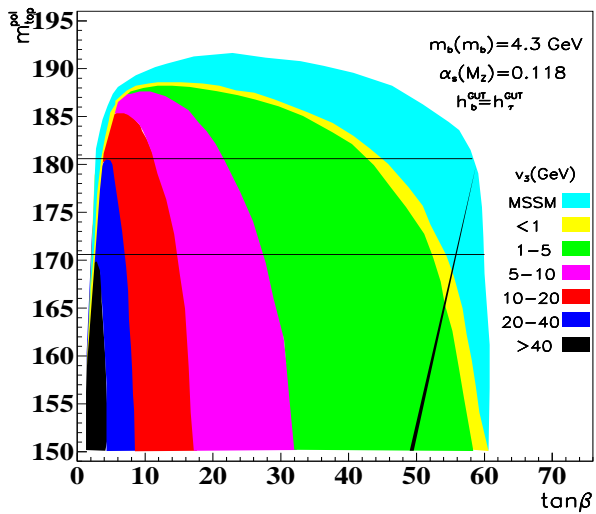,height=7.0truecm,width=9truecm}}}
\caption{Pole top quark mass as a function of $\tan\beta$ for different
values of the R--Parity violating parameter $|v_3|$. Bottom quark and tau
lepton Yukawa couplings are unified at $M_{GUT}$. The horizontal lines 
correspond to the $1\sigma$ experimental determination of $m_t$. Points 
with $t-b-\tau$ unification are concentrated in the diagonal line at high 
values of $\tan\beta$.}
\label{aretop}
\end{figure}
These differences with the MSSM have profound consequences on Yukawa 
unification as shown in Fig.~\ref{aretop}. In this figure we observe 
that bottom--tau Yukawa unification can be achieved at any value of 
$\tan\beta$ by chosing appropriately the value of $v_3$ \cite{YukUnif}.
The plot in Fig.~\ref{aretop} is made with a scan over parameter space
such that points which satisfy $h_b(M_{GUT})=h_{\tau}(M_{GUT})$ within
$1\%$ are kept, where $M_{GUT}$ is the gauge coupling unification scale.
Each selected point is placed in one of the regions of Fig.~\ref{aretop}
according to its $|v_3|$ value. The diagonal band at high values of 
$\tan\beta$ corresponds to points where top-bottom-tau unification is
achieved \cite{YukUnif}.

In summary, it is shown that BRpV is the simplest extension of the MSSM
which introduce R--Parity violation. This model can be successfully 
embedded into Supergravity models with universality of scalar, gaugino, 
bilinear and trilinear soft mass parameters. In this case, the induced
tau neutrino mass is radiatively generated and, therefore, naturally
small. It is shown that the BRpV parameters $\epsilon_3$ and $v_3$ do
not need to be small, in fact they can be easily of the order of $m_Z$.
In addition, in this model the CP--even Higgs bosons couple with the
tau sneutrino field, and the effect on the mass of the lightest Higgs
is to lower it compared to the MSSM. Finally, BRpV changes the relation
between the Yukawa couplings and the masses of the top and bottom quarks 
and the tau lepton. As a consequence, bottom-tau Yukawa unification can
be achieved at any value of the parameter $\tan\beta$ provided we choose
appropriately the value of the sneutrino vev $v_3$. Top-bottom-tau
unification is achieved in a slightly wider region at high $\tan\beta$.
We would like to stress the fact that, even in the unlikely limit where 
the tan neutrino is massless with $\epsilon_3\ne0$ (if $v'_3=0$, obtained 
when there is universality of soft mass parameters {\it{at the weak 
scale}}, which is not natural) R--Parity is not conserved, and even 
though the neutralinos decouple from the tau neutrino, the lightest 
neutralino decays for example to $b\overline b \nu_{\tau}$ through an 
intermediate sbottom.

\section*{Acknowledgments:} 

The author is indebted to his collaborators A. Akeroyd, 
J. Ferrandis, M.A. Garcia--Jare\~no, A. Joshipura, J.C. Rom\~ao, 
and J.W.F. Valle for their contribution to the work presented here.
The author was supported by a postdoctoral grant from Ministerio de 
Educaci\'on y Ciencias, by DGICYT grant PB95-1077 and by the EEC 
under the TMR contract ERBFMRX-CT96-0090.


\begin{thebibliography}{99}

\bibitem{MSSM}
H.P. Nilles, {\sl Phys. Rep.} {\bf 110}, 1 (1984); 
H.E. Haber and G.L. Kane, {\sl Phys. Rep.} {\bf 117}, 75 (1985); 
R. Barbieri, {\sl Riv. Nuovo Cimento} {\bf 11}, 1 (1988). 

\bibitem{gaugeUnif}
U. Amaldi, W. de Boer, and H. Furstenau, {\sl Phys. Lett. B} {\bf 260},
447 (1991); J. Ellis, S. Kelley, and D.V. Nanopoulos, {\sl Phys. Lett. B}
{\bf 260}, 131 (1991); P. Langacker and M. Luo, {\sl Phys. Rev. D}
{\bf 44}, 817 (1991); C. Giunti, C.W. Kim and U.W. Lee, {\sl Mod. Phys. 
Lett.} {\bf A6}, 1745 (1991).

\bibitem{gaugUnifRecent}
P. Langacker and N. Polonsky, {\sl Phys. Rev. D}{\bf 47}, 4028 (1993); 
P.H. Chankowski, Z. Pluciennik, and S. Pokorski, {\sl Nucl. Phys. B} 
{\bf 439}, 23 (1995);
P.H. Chankowski, Z. Pluciennik, S. Pokorski, and C.E. Vayonakis, 
{\sl Phys. Lett. B}{\bf 358}, 264 (1995).

\bibitem{SUSY}
Yu.A. Gol'fand and E.P. Likhtman, {\sl JETP Lett.}{\bf 13}, 323 (1971); 
D.V. Volkov and V.P. Akulov, {\sl JETP Lett.} {\bf 16}, 438 (1972);
J. Wess and B. Zumino, {\sl Nucl. Phys.} {\bf B70}, 39 (1974).

\bibitem{e3others}
F. de Campos, M.A. Garc{\'\i}a-Jare\~no, A.S. Joshipura, J. Rosiek, 
and J.W.F. Valle, {\sl Nucl. Phys.} {\bf B451}, 3 (1995); 
A.S. Joshipura and M. Nowakowski, {\sl Phys. Rev. D} {\bf 51}, 2421 (1995);
T. Banks, Y. Grossman, E. Nardi, and Y. Nir, {\sl Phys. Rev. D}
{\bf 52}, 5319 (1995);
F. Vissani and A.Yu. Smirnov, {\sl Nucl.Phys.} {\bf B460}, 37 (1996); 
R. Hempfling, {\sl Nucl. Phys.} {\bf B478}, 3 (1996); 
F.M. Borzumati, Y. Grossman, E. Nardi, Y. Nir, {\sl Phys. Lett. B}
{\bf 384}, 123 (1996);
H.P. Nilles and N. Polonsky, {\sl Nucl. Phys.} {\bf B484}, 33 (1997); 
B. de Carlos, P.L. White, {\sl Phys. Rev. D} {\bf 55}, 4222 (1997); 
E. Nardi, {\sl Phys. Rev. D} {\bf 55}, 5772 (1997);
S. Roy and B. Mukhopadhyaya, {\sl Phys. Rev. D} {\bf 55}, 7020 (1997);
A. Faessler, S. Kovalenko, F. Simkovic, hep-ph/9712535;
M. Carena, S. Pokorski, and C.E.M. Wagner, hep-ph/9801251; 
M.E. G\'omez and K. Tamvakis, hep-ph/9801348.
 
\bibitem{BRpVtalk}
M.A. D\'\i az, hep-ph/9711435; M.A. D\'\i az, hep-ph/9712213;
J.W.F. Valle, hep-ph/9712277; J.C. Rom\~ao, hep-ph/9712362.

\bibitem{SRpSB}
A. Masiero and J.W.F. Valle, {\sl Phys. Lett. }{\bf B251}, 273 (1990);
J.C. Rom\~ao, A. Ioannissyan and J.W.F. Valle,
{\sl Phys. Rev. D}{\bf 55}, 427 (1997).

\bibitem{HallSuzuki}
L. Hall and M. Suzuki, {\sl Nucl.Phys.} {\bf B231}, 419 (1984).

\bibitem{Basis}
M.A. D\'\i az, A.S. Joshipura, and J.W.F. Valle, in preparation.
   
\bibitem{epsrad} 
M.A. D\'\i az, J.C. Rom\~ao, and J.W.F. Valle, hep-ph/9706315.

\bibitem{ChaStau}
A. Akeroyd, M.A. D\'\i az, J. Ferrandis, M.A. Garcia--Jare\~no, and 
J.W.F. Valle, hep-ph/9707395; M.A. D\'\i az, hep-ph/9710233; 
J. Ferrandis, hep-ph/9802275.

\bibitem{mhRadCorr}
H.E. Haber and R. Hempfling,{\sl Phys. Rev. Lett.} {\bf 66}, 1815 (1991);
A. Yamada, {\sl Phys. Lett. B} {\bf 263}, 233 (1991); Y. Okada, M. 
Yamaguchi and T. Yanagida, {\sl Phys. Lett. B} {\bf 262}, 54 (1991);
J. Ellis, G. Ridolfi and F. Zwirner, {\sl Phys. Lett. B}{\bf 257}, 83 
(1991); R. Barbieri, M. Frigeni, F. Caravaglios, {\sl Phys. Lett. B} 
{\bf 258}, 167 (1991); J.L. Lopez and D.V. Nanopoulos, 
{\sl Phys. Lett. B} {\bf 266}, 397 (1991); A. Brignole, 
{\sl Phys. Lett. B} {\bf 281}, 284 (1992); D.M. Pierce, A. Papadopoulos, 
and S. Johnson, {\sl Phys. Rev. Lett.} {\bf 68}, 3678 (1992); 
M.A. D\'\i az and H.E. Haber, {\sl Phys. Rev. D} {\bf 46}, 3086 (1992).

\bibitem{YukUnif}
M.A. D\'\i az, J. Ferrandis, J.C. Rom\~ao, and J.W.F. Valle, 
hep-ph/9801391.

\end{thebibliography}
\end{document}